\documentclass[journal]{IEEEtran}
\usepackage{amsmath,amsfonts,amssymb}
\usepackage{algorithm}
\usepackage{url}
\usepackage{graphicx}
\usepackage{cite}
\usepackage{algpseudocode}
\usepackage{xcolor}
\usepackage[caption=false,font=normalsize,labelfont=sf,textfont=sf]{subfig}
\usepackage{textcomp}
\usepackage{stfloats}

\begin{document}

\title{Signed Powers-of-Two Greedy Vector Approximation \\ of Zadoff-Chu Sequences}

\author{Luiz F. da S. Coelho,~\IEEEmembership{Member,~IEEE,}
\and Paulo S. R. Diniz,~\IEEEmembership{Life Fellow,~IEEE,}
\and Didier Le Ruyet,~\IEEEmembership{Senior Member,~IEEE}
        % <-this % stops a space
\thanks{Luiz F. da S. Coelho is a PhD candidate at PEE-COPPE/UFRJ, Brazil, and CEDRIC-CNAM, France. Paulo S. R. Diniz is with PEE-COPPE/UFRJ, Brazil. Didier Le Ruyet is with CEDRIC-CNAM, France.

This work was supported in part by the Coordenação de Aperfeiçoamento de Pessoal de Nível Superior--Brasil (CAPES)--Finance Code 001, by FAPERJ (the Rio de Janeiro State research council), and by the National Institute of Science and Technology (INCT) STREAM, the Brazilian National Council for Scientific and Technological Development (CNPq), under Grant 409179/2024-8. This work was partially supported by the ANR under the France 2030 program, grant ANR-22-PEFT-0004.}
}

\markboth{Journal of \LaTeX\ Class Files,~Vol.~14, No.~8, August~2021}%
{Shell \MakeLowercase{\textit{et al.}}: A Sample Article Using IEEEtran.cls for IEEE Journals}

\IEEEpubid{0000--0000/00\$00.00~\copyright~2021 IEEE}

\maketitle

\begin{abstract}
This paper explores the use of sums of signed powers-of-two (SOPOT) for a representation suitable for multiplierless applications of ZC sequences, where multiplications are reduced to a prescribed number of shift-and-add operations. The SOPOT representation is achieved using a greedy algorithm that outperforms other approaches for multiplierless approximations, such as CORDIC and CSD. The proposed algorithm is based on the matching pursuits using generalized bit planes (MPGBP), and the proposed modification enables robustness to hyperparameter tuning. The resulting sequences retain, to some extent, the good autocorrelation properties of ZC, as shown in a matched-filter detection simulation. This corroborates to their use in various practical applications across mainstream cellular systems and devices requiring moderate computational cost.
\end{abstract}

\vspace{-1ex}
\section{Introduction}
Zadoff-Chu (ZC) sequences are widely used in LTE and 5G for initial downlink synchronization, random access, uplink control information, and uplink pilot signals~\cite {andrews2023primerzadoffchusequences,omri2019synchronization,sesia2011lte,tuninato20245g}. The ZC sequences are also used for carrier-frequency offset (CFO) estimation and preamble detection in OFDM systems. The ZC is a complex sequence with ideal cyclic autocorrelation, with a single nonzero value at the zero-lag entry. Given their frequent use across multiple wireless communication systems, it is valuable to verify whether efficient, practical implementations of ZC sequences exist that incur low computational cost. The latter is desirable for communication and sensing nodes that rely on battery power.

The literature proposes several efficient implementations of ZC sequences. Storage is the first limitation; complex sequences require substantial memory to encode at full precision (\texttt{float32}) and can benefit from low-precision representations~\cite{krondorf2022icarus,mansour2013hardware,de2014efficient}. Moreover, to exploit its good autocorrelation properties, we use cross-correlation, which can be computationally demanding in full precision. Hence, using a smart quantization method to obtain an appropriate numerical representation is advisable~\cite {mansour2013hardware,de2014efficient}.

To address the high memory requirement of ZC sequences, \cite{krondorf2022icarus} approximates the sequences using phase shift keying (PSK) modulation, specifically 8-PSK. Although this approach enables 3-bit per-element quantization, it ignores the computational cost of the subsequent cross-correlation operation. Therefore, to reduce the computational cost of complex multiplications, the coordinate rotation digital computer (CORDIC) relies on iterative phase-shift operations~\cite {mansour2013hardware,de2014efficient}.

We propose using sums-of-signed-powers-of-two (SOPOT) representation. SOPOT enables efficient shift-and-add operators instead of a conventional multiplication module. In SOPOT, numbers are represented using signed powers of two (SPTs), extending the bit plane into a generalized, or signed, bit plane that allows \textit{negative bits}. This expansion sparsifies the bit plane, enabling fewer active bits. The number of active bits defines the number of shift-and-add operations in the multiplication; hence, the interest in a sparse representation.

The matching pursuits with generalized bit planes (MPGBP) algorithm~\cite{da2014fir} proposes an SOPOT representation of finite impulse response (FIR) filters with a computationally efficient implementation. Another SOPOT method is the canonical signed digit (CSD), which converts a fixed-point representation into a sparse signed-digit representation. The MPGBP algorithm has been successfully used in multiplierless adaptive filters~\cite{da2021adaptive} and in post-training quantization of a multilayer perceptron deep neural network~\cite{coelho2025multiplierless}, outperforming CSD.
\IEEEpubidadjcol

This work expands our understanding of the MPGBP algorithm and advances its application to quantization. We propose a modified MPGBP algorithm that better controls the number of SPTs, subject to a maximum bit plane depth, and explore its use in quantizing ZC sequences for multiplierless computation. Our solution outperforms state-of-the-art quantization methods while maintaining low computational complexity.

Section~\ref{sec:sopot} details the SOPOT approximation method. Section~\ref{sec:zc-seq} briefly introduces the ZC sequence and the quantization framework used for the simulations. Section~\ref{sec:results} presents and discusses the numerical simulation results. Section~\ref{sec:conclusion} concludes the work.
\vspace{-2ex}
\section{SOPOT Approximation} \label{sec:sopot}
\vspace{-1ex}
Let ${\bf x}$ be a $N$-length vector representing a real-valued signal, where ${\bf x} \in \mathbb{R}^N$, composed as \vspace{-1ex}
\begin{align}
    {\bf x} = \begin{bmatrix}
        x(0) & x(1) & \cdots & x(N-1)
    \end{bmatrix}^T,
    \vspace{-1ex}
\end{align}
where $[\cdot]^T$ returns the transpose. Let the inner product of two $N$-length vectors, ${\bf u} \in \mathbb{R}^N$, ${\bf v} \in \mathbb{R}^N$, be defined as \vspace{-1ex}
\begin{align}
    a = \left\langle {\bf u}, {\bf v} \right\rangle = {\bf u}^T {\bf v} = \sum_{n=0}^{N-1} u(n) v(n),
\end{align}
and, for ${\bf v}$ as an unit-norm vector, i.e., $||{\bf v}||_2=1$, where $||\cdot||_2$ is the Euclidean norm, the orthogonal projection of ${\bf u}$ in the vector ${\bf v}$ is defined as
\begin{align}
    a{\bf v} = \left\langle {\bf u}, {\bf v} \right\rangle {\bf v} = \left({\bf u}^T{\bf v}\right){\bf v},
\end{align}
where, in this context, the inner product, $a \in \mathbb{R}$, is the projection coefficient.

Following the proposition in~\cite{da2014fir}, let $\mathcal{M} = \{-1, 0, 1\}$, and $\mathcal{D} = \{\boldsymbol{\phi}_p\}_{p\in\Gamma}$ be a dictionary of $P$ $N$-length vectors composed of elements of $\mathcal{M}$, i.e., $\boldsymbol{\phi}_p \in \mathcal{M}^N$ and $\Gamma = \{0, 1, \dots, P-1\}$ is the set with indexes for the dictionary vectors; moreover, the dictionary vectors are constrained by the number of nonzero elements, $L$, where $L=||\boldsymbol{\phi}_p||_0, \forall p\in\Gamma$, with $||\cdot||_0$ as the pseudo-norm $\ell_0$. Let us consider the signal ${\bf x} \in \mathbb{R}^N$, such that $||{\bf x}||_{\infty} \leq 1$, without loss of generality, ${\bf x}$ can be approximated using the $M$-term SOPOT as follows
\begin{align}
    {\bf x} \approx \hat{\bf x}_M = \sum_{m=0}^{M-1} Q(a(p_m)) \boldsymbol{\phi}_{p_m} = \sum_{m=0}^{M-1} 2^{-k_m} \boldsymbol{\phi}_{p_m}, \label{eq:m-term}
\end{align}
where $M = \{M \in \mathbb{Z} : M < P\}$, $\hat{\bf x}_M \in \mathbb{R}^N$ is the $M$-term approximation of ${\bf x}$, $Q(\cdot)$ is a quantization function defined in Appendix~\ref{app:qtz-fun}, $\boldsymbol{\phi}_{p_m}$ is the $m$-th vector selected from $\mathcal{D}$, with $p_m \in \Gamma$ chosen following the criterion described in Appendix~\ref{app:vect-select}, $a(p_m) \in \mathbb{R}$ is the projection coefficient for $\boldsymbol{\phi}_{p_m}$, and $k_m \in \mathbb{Z}$ is the signed bit plane depth, selected by $Q(\cdot)$.

The approximation is achieved based on a matching pursuits (MP) algorithm, with residues estimated iteratively as follows
\begin{align}
    {\bf r}_{m+1} &= {\bf r}_m - \left\langle {\bf r}_m, \frac{\boldsymbol{\phi}_{p_m}}{\sqrt{L}} \right\rangle \frac{\boldsymbol{\phi}_{p_m}}{\sqrt{L}} \nonumber \\
    &= {\bf r}_m - \frac{{\bf r}_m^T\boldsymbol{\phi}_{p_m}}{L} \boldsymbol{\phi}_{p_m} = {\bf r}_m - a(p_m) \boldsymbol{\phi}_{p_m},
\end{align}
where ${\bf r}_m \in \mathbb{R}^N$ is the $m$-th residue vector, and the algorithm should start at ${\bf r}_0 = {\bf x}$. The algorithm is further approximated using the quantization of the projection coefficient,
\begin{align}
    {\bf r}_{m+1} = {\bf r}_m - Q(a(p_m))\boldsymbol{\phi}_{p_m}.
\end{align}
Although the quantization process is nonlinear, \cite{da2014fir} proves that the algorithm still converges.

The sum in \eqref{eq:m-term} numerically approximates ${\bf x}$ using $M$ elements of the finite dictionary $\mathcal{D}$ allocated across multiple signed bit planes with varying depths $k_m$. There are no constraints on assigning a non-zero to an already non-zero element in the bit plane, and this is not problematic for the approximation. However, for hardware implementation, the key elements are the active terms in signed bit planes, or the number of SPTs. The seminal work~\cite{da2014fir} proposes a common term reduction (CTR) algorithm as a post-processing phase to extract such terms. Their proposition uses $L_{\textrm{max}}$ as an upper bound to the SPT count.

Nonetheless, we propose an online CTR approach that gives the algorithm better control over the number of SPTs given a budget. We begin by defining the vector approximation in the form of bit planes. Let $\mathcal{K}$ be the set of assigned values for $k_m$. We define $k_{\textrm{max}} \triangleq \sup(\mathcal{K})$ as the maximum bit plane depth, and we know, from $||{\bf x}||_\infty \leq 1$, that $\inf(\mathcal{K}) = 0$; let $\boldsymbol{\Phi}_m$ be the signed bit plane matrix
\begin{align}
    \boldsymbol{\Phi}_m &= \begin{bmatrix}
        \bar{\boldsymbol{\varphi}}(0) & \bar{\boldsymbol{\varphi}}(1) & \cdots & \bar{\boldsymbol{\varphi}}(N-1))
    \end{bmatrix}^T,
\end{align}
where $\bar{\boldsymbol{\varphi}}(n) \in \mathcal{M}^{k_{\textrm{max}}+1}$ is the vector representing the $n$-th row of $\boldsymbol{\Phi}_m \in \mathcal{M}^{N \times (k_{\textrm{max}}+1)}$ and defined as
\begin{align}
    \bar{\boldsymbol{\varphi}}(n) = \begin{bmatrix}
        \varphi_n(0) & \varphi_n(1) & \cdots & \varphi_n(k_{\textrm{max}})
    \end{bmatrix}^T,
\end{align}
with $\varphi_n(k) \in \mathcal{M}$ as the $k$-th element of the $n$-th row of $\boldsymbol{\Phi}_m$. Moreover, let us define the signed bit plane vector, ${\bf b}$, as
\begin{align}
    {\bf b} \triangleq \begin{bmatrix}
        2^0 & 2^{-1} & \cdots & 2^{-k_{\textrm{max}}}
    \end{bmatrix}^T,
\end{align}
with ${\bf b} \in \mathbb{R}^{k_{\textrm{max}}+1}$. Consider that $\boldsymbol{\Phi}_m$ accumulates all the $m$ $\boldsymbol{\phi}_{p_m}$ vectors at their respective $k_m$ columns; there exists a set of rules applied whenever $\boldsymbol{\Phi}_m$ incorporates a new dictionary vector $\boldsymbol{\phi}_{p_m}$ that ensures $\boldsymbol{\Phi}_m \in \mathcal{M}^{N \times (k_{\textrm{max}}+1)}$ and, at the final ($M$-th) iteration \vspace{-2ex}
\begin{align}
    \hat{\bf x}_M = \boldsymbol{\Phi}_M {\bf b} = \sum_{m=0}^{M-1} 2^{-k_m} \boldsymbol{\phi}_{p_m}.
    \vspace{-1ex}
\end{align}

Algorithm~\ref{alg:ctr} is a procedure for achieving a sparse representation of the matrix $\boldsymbol{\Phi}_M$. It ensures that the budget, $L_{\textrm{max}}$, is always attained if the maximum depth for the bit planes, $k_{\textrm{max}}$, is sufficiently large.
In brief, lines 3 to 5 check and resolve whether an SPT is allocated twice in the same placement; moreover, lines 6 to 13 check and resolve whether two neighboring planes have SPTs. Assuming $||{\bf x}||_\infty \leq 1$ ensures that $k_m \leq 0$ and that two SPTs are not allocated twice at $k_m=0$.

\begin{algorithm}[!t]
    \caption{Online CTR}\label{alg:ctr}
    \begin{algorithmic}[1]
        \Procedure{add\_spt}{$\bar{\boldsymbol{\varphi}}(n), k_m, \phi_{p_m}(n)$}
        \State{$\varphi_n(k) \gets \varphi_n(k) + \phi_{p_m}(n)$}
        \If{$|\varphi_n(k)| = 2$}
        \State{$\varphi_n(k) \gets 0$}
        \State{$\bar{\boldsymbol{\varphi}}(n) \gets$ {\Call{add\_spt}{$\bar{\boldsymbol{\varphi}}(n), k_m-1, \phi_{p_m}(n)$}}}
        \EndIf
        \If{$k_m>0 ~\&~ \varphi_n(k)+\varphi_n(k-1)\pmod{2} = 0$}
        \If{$\varphi_n(k)+\varphi_n(k-1) = 0$}
        \State{$\varphi_n(k) \gets \varphi_n(k-1)$}
        \State{$\varphi_n(k-1) \gets 0$}
        \Else
        \State{$\bar{\boldsymbol{\varphi}}(n)\gets$ {\Call{add\_spt}{$\bar{\boldsymbol{\varphi}}(n), k_m-2, \phi_{p_m}(n)$}}}
        \State{$\varphi_n(k-1) \gets 0$}
        \State{$\varphi_n(k) \gets -\varphi_n(k)$}
        \EndIf
        \EndIf
        \State\Return{$\bar{\boldsymbol{\varphi}}(n)$}
        \EndProcedure
    \end{algorithmic}
\end{algorithm}
\begin{algorithm}[!t]
    \caption{Modified MPGBP for SOPOT approximation}\label{alg:mpgbp}
    \begin{algorithmic}[1]
        \Require{${\bf x}$, $L$, $L_{\textrm{max}}$, and $k_{\textrm{max}}$, $\epsilon$}
        \Ensure{$\hat{\bf x}_m$, $\boldsymbol{\Phi}_m$}
        \State $m \gets 0$, ${\bf r}_0 \gets {\bf x}$, $\hat{\bf x}_0 \gets {\bf 0}$, $\boldsymbol{\Phi}_0 \gets {\bf 0}$
        \While{$||\boldsymbol{\Phi}_m||_F^2 < L_{\textrm{max}}$ \& $\frac{{\bf r}_m^T{\bf r}_m}{N} > \epsilon$}
        \State{$L_m \gets \min(L, L_\textrm{max}-||\boldsymbol{\Phi}_m||_F^2)$}
        \State Sort indexes of ${\bf r}_m$ in decreasing order of the absolute value; store the first $L_m$ indexes in $\mathcal{L}_m$
        \State $\boldsymbol{\phi}_{p_m} \gets {\bf 0}$
        \For{$l \in \mathcal{L}_m$}
        \State $\phi_{p_m}(l) \gets \textrm{sign}(r_m(l))$
        \EndFor
        \State $k_m \gets \left\lceil -\log_2 \left( \frac{4}{3L_{m}} {\bf r}_m^T\boldsymbol{\phi}_{p_m} \right) \right\rceil$
        \If{$k_m > k_{\textrm{max}}$}
        \State{\textbf{break}}
        \EndIf
        \State $\hat{\bf x}_{m+1} \gets \hat{\bf x}_m + 2^{-k_m} \boldsymbol{\phi}_{p_m}$
        \State ${\bf r}_{m+1} \gets {\bf r}_m - 2^{-k_m} \boldsymbol{\phi}_{p_m}$
        \State $m \gets m+1$
        \For{$n = 0, 1 , \dots, N-1$}
        \State{$\bar{\boldsymbol{\varphi}}(n) \gets$ \Call{add\_spt}{$\bar{\boldsymbol{\varphi}}(n), k_m, \phi_{p_m}(n)$}}
        \EndFor
        \EndWhile
    \end{algorithmic}
\end{algorithm}

Algorithm~\ref{alg:mpgbp} presents the modified MPGBP, which enables achieving better control over the number of SPTs in $\boldsymbol{\Phi}_M$. $L_{\textrm{max}} \geq ||\boldsymbol{\Phi}_M||_F^2$, where $||\cdot||_F$ is the Frobenius norm; $L_\textrm{max}$ can be used as a stopping criterion. Another stopping criterion is the approximation mean squared error (MSE), calculated as
\begin{align}
    \textrm{MSE}_m = \frac{||\hat{\bf x}_m - {\bf x}||_2^2}{N} = \frac{{\bf r}_m^T{\bf r}_m}{N},
\end{align}
where $\textrm{MSE}_m$ is the MSE at the $m$-th iteration, limited by a threshold $\epsilon$. Moreover, $k_\textrm{max}$ is another proposed stopping criterion; it profits from the fact that $k_m$ is monotonically non-decreasing over iterations. $k_{\textrm{max}}$ is also important in the quantization aspect of the algorithm, where it limits the number of possible values for the sequence; see~\ref{subsec:mpgbp}.
\vspace{-2ex}
\section{The Zadoff-Chu Sequence Quantization and Multiplierless Implementation} \label{sec:zc-seq}
The ZC sequence, proposed in~\cite{chu1972polyphase} as an extension of the work developed in~\cite{frank1962phase}, for generating sequences of arbitrary length. The ZC sequence is defined as
\begin{align}
    x(n) = \begin{cases}
        e^{j\frac{K \pi n^2}{N}}, &\quad N \textrm{ even}\\
            e^{j\frac{K\pi n(n+1)}{N}}, &\quad N \textrm{ odd}
    \end{cases}, \label{eq:zc}
\end{align}
where $N \in \mathbb{Z}$ is the sequence length, $n=0, 1, \dots, N-1$ indexes the sequence vector ${\bf x} \in \mathbb{C}^N$, and $K \in \mathbb{Z}$ is the root of the sequence, with $K<N$ as a relatively prime to $N$.

The ZC sequences have constant (unit) amplitude, with an ideal cyclic autocorrelation calculated as \vspace{-1ex}
\begin{align}
    r_{xx}(\tau) = \sum_{n=0}^{N-1} x(n) x^\ast ((n+\tau)\pmod N) = N \delta(\tau),
    \vspace{-1ex}
\end{align}
where $r_{xx}(\tau)$ is the discrete periodic autocorrelation function of $x$ at the delay $\tau$, and $\delta(\tau)$ is the Kronecker delta. Moreover, the cross-correlation between two ZC sequences of the same length is constant. It equals $\sqrt{N}$ if the absolute difference between the roots, $|K_1 - K_2|$, is relatively prime to $N$, where $K_1$ and $K_2$ are the roots for each sequence. This is the minimum value for sequences with ideal autocorrelation~\cite{sesia2011lte}.
\vspace{-2ex}
\subsection{Sequence Quantization}
\subsubsection{MPGBP} \label{subsec:mpgbp}
We use Algorithm~\ref{alg:mpgbp} to quantize the ZC sequences' in-phase and quadrature parts separately. In the MPGBP-quantized sequences, the number of possible values is $2^{k_{\textrm{max}}+1}+1$, ranging in the interval $[-1,1]$ and stepped by $2^{-k_{\textrm{max}}}$, for each component. The budget, $L_{\textrm{max}}$, defines a bottleneck on the maximum number of nonzero elements in the signed bit plane matrix. Small values of $L_{\textrm{max}}$ reduce the precision of the approximation, reducing the sequence's entropy and increasing the MSE. Limiting $k_{\textrm{max}}$ constrains the maximum bit shift, which may be constrained by the hardware design; however, small values yield coarser approximations. $L$ varies the number of vector elements processed per iteration; large values may induce high MSE, while small values increase the number of iterations before convergence. In~\cite{da2014fir}, $L=\left\lfloor\sqrt{N}\right\rfloor$ offers a good compromise between MSE degradation and computational time. $\epsilon$ is introduced as a ceiling for a good enough approximation error, and early stops the algorithm without allocating all resources.

\subsubsection{CSD}
We achieve the CSD representation via a post-processing step applied to the 2's-complement representation. In this work, a $b$-bit CSD means that the real number is quantized using a $b$-bit word length. We consider $b-2$ bits as decimal digits in our conversion, with one bit for the sign and one bit to represent the integers.
\subsubsection{PSK}
In the PSK approximation~\cite{krondorf2022icarus}, the sequence elements map to the PSK constellation, enabling direct encoding; for example, 8-PSK uses 3 bits per sequence element.
\vspace{-3ex}
\subsection{Sequence Encoding}
As multiple ZC sequences might be stored and used by the same device, the bit representation for storing the sequences is computed using the set of quantized values across all sequences. For CSD and PSK, the required number of bits comes directly from the quantization; however, MPGBP requires an extra step. Let $\Lambda$ be the set of complex values from all quantized sequences, and the set size is defined as $|\Lambda|$; the number of bits, $R$, necessary to store a single sequence is calculated as
\begin{align}
    R = N \left\lceil \log_2 \left( |\Lambda | \right) \right\rceil,
\end{align}
where $\lceil\cdot\rceil$ is the ceiling operator. Although an encoding that benefits from sequence sparsity exists (e.g., Huffman encoding), the proposed encoding adds no processing overhead. We disregard instruction-wise considerations, as both CORDIC and SOPOT depend on the hardware choice.

\vspace{-2ex}
\subsection{Sequence Implementation}
Complex numbers are multiplied using 4 real-valued multiplications and two additions. SOPOT representation enables shift-and-add operations for each real-valued multiplication. A SOPOT using $l$ SPTs represents $l$ shifts and $l-1$ additions. Nonetheless, CORDIC enables iterative phase rotation with each iteration consisting of 2 shift-and-add operations~\cite{andraka1998survey}.

In terms of computational complexity, shifts are not as significant; therefore, we count only the number of additions, $A_n$, for the multiplication by the $n$-th sequence element. For the PSK+CORDIC, $A_n=2\theta$, where $\theta$ is the number of iterations, and for the SOPOT-based quantization, we have $A_n = \max( 2(l_{n,\textrm{I}} + l_{n,\textrm{Q}}) - 2, 0)$, where $l_{n,\textrm{I}}$ and $l_{n,\textrm{Q}}$ are the number of SPTs in the in-phase and quadrature parts of the $n$-th element, respectively. Let $\Theta$ be the set of all $n$ where $A_n \neq 0$; the total number of additions per inner product is
\begin{align}
    A = \sum_{n \in \Theta} A_n + |\Theta|-1,
\end{align}
where $|\Theta|$ is the set size.
\vspace{-1ex}
\section{Results}\label{sec:results}
Considering 5G-compliant ZC sequences~\cite{3gpp_38211}, we use 64 ZC sequences of length $N=839$, and roots defined by the logical index $\{0, 1, \dots, 63\}$. We compare the quantization methods in terms of the complexity of implementing an inner product, the core operation in cross-correlation. We consider MPGBP approximations using the following fixed parameters, $k_{\textrm{max}} = 6$, $L=\left\lfloor\sqrt{N}\right\rfloor$, and $\epsilon=10^{-6}$.

Table~\ref{tab:qtz-results} presents the results for each quantization method. It shows the average complexity, $\bar{A}$, and average MSE with standard deviation. It also presents the number of bits, $R$, per sequence. Notably, MPGBP offers more flexibility than the other solutions, enabling fewer additions for a good MSE, e.g., $\bar{A} = 1883$ for an average MSE of $-18.43~\textrm{dB}$.
\begin{table}[!t]
    \centering
    \caption{Average complexity, $\bar{A}$, and MSE, with standard deviation and number of bits, $R$, for the considered ZC sequences.}
    \begin{tabular}{c||c|c|c}
        \textbf{Quantization} & $\bar{A}$ & MSE (dB) & $R$ \\ \hline\hline
        None & - & - & $13424$ \\
        $16$-PSK $(\theta=4)$ & $6711\pm0$ & $-17.80\pm0.16$ & $3356$ \\
        64-PSK $(\theta=6)$ & $10067\pm0$ & $-29.58\pm0.19$ & $5034$ \\
        256-PSK $(\theta = 8)$ & $13423\pm0$ & $-41.59\pm0.17$ & $6712$ \\
        4-bit CSD & $2675\pm 17$ & $-12.73\pm0.05$ & $6712$ \\
        6-bit CSD & $4621\pm51$ & $-25.25\pm0.09$ & $10068$ \\
        8-bit CSD & $6563\pm55$ & $-37.53\pm0.10$ & $13424$ \\
        MPGBP ($L_\textrm{max} = 84$) & $167\pm0$ & $-0.96\pm0.00$ & $2517$ \\
        MPGBP ($L_\textrm{max} = 420$) & $839\pm0$ & $-7.05\pm0.06$ & $4195$ \\
        MPGBP ($L_\textrm{max} = 839$) & $1883\pm8$ & $-18.43\pm0.20$ & $5034$
    \end{tabular}
    \label{tab:qtz-results}
    \vspace{-2ex}
\end{table}

\begin{figure}[!t]
    \centering
    \subfloat[]{\includegraphics[width=.46\linewidth]{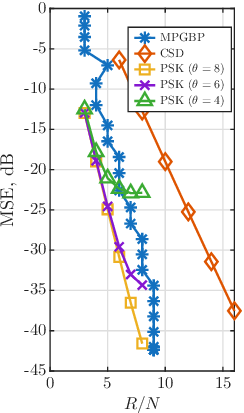}
        \label{fig:mse-r}}
    \hfill
    \subfloat[]{\includegraphics[width=.485\linewidth]{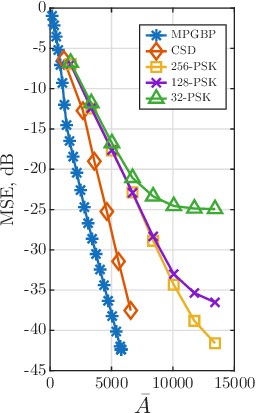}
        \label{fig:mse-l}}
    \caption{Average MSE for (a) number of bits per element, $R/N$, with PSK varying from $8$- to $256$-PSK, and (b) average number of additions, $\bar{A}$, with PSK varying $\theta$ from $1$ to $8$. MPGBP varies $L_{\textrm{max}}$ from 84 to 2098 (steps of 84), and CSD uses 3 to 8 bits.}
    \label{fig:mse}
    \vspace{-2.5ex}
\end{figure}

Although the PSK offers advantages in storing sequences, it requires more additions than the MPGBP to achieve a similar MSE. This behavior is confirmed in Fig.~\ref{fig:mse}, where we present the average MSE in terms of (a) bits per element, $R/N$, and (b) the average number of additions, $\bar{A}$. Fig.~\ref{fig:mse-r} shows that the MPGBP is outperformed only by the PSK with comparable storage size. However, Fig.~\ref{fig:mse-l} shows that the MPGBP outperforms both solutions when we compare MSE for the number of additions. For $\bar{A} \approx 3500$, $L_{\textrm{max}} = 1175$, we have gains of $7~\textrm{dB}$ when compared to the 5-bit CSD and $15~\textrm{dB}$ against the PSK $(\theta=2)$.

\begin{figure}[!t]
    \centering
    \includegraphics[width=\linewidth]{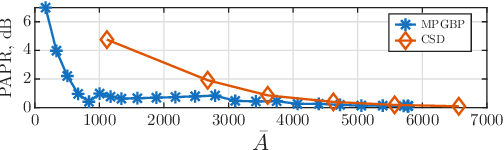}
    \caption{Average PAPR in terms of the average number of additions, $\bar{A}$, for MPGBP using $L_{\textrm{max}}$ varying from $84$ to $2098$ (steps of $84$) and CSD using $3$ to $8$ bits.}
    \label{fig:papr}
    \vspace{-2.5ex}
\end{figure}

Another important aspect of ZC sequences is the peak-to-average power ratio (PAPR)~\cite{sesia2011lte}, which is constant at $0~\textrm{dB}$ for non-quantized sequences. The PSK approximations also achieve $0~\textrm{dB}$, and the results for the MPGBP and CSD approximations are presented in Fig.~\ref{fig:papr}. Higher-precision approximations yield lower PAPR.
When $L_{\textrm{max}}=420$, we approach a local minimum for PAPR at $0.41~\textrm{dB}$ with $\bar{A}=839$ and $R/N = 5$. This is only outperformed when $\bar{A} \geq 4000$ and $R/N \geq 8$, with a PAPR of $0.26~\textrm{dB}$. In our experiments, when comparing PAPR in terms of bits per element, MPGBP outperforms CSD in all tested scenarios.

\begin{figure}[!t]
    \centering
    \subfloat[]{\includegraphics[width=0.45\linewidth]{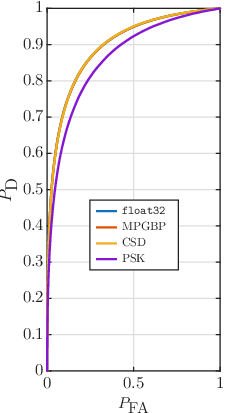}\label{fig:roc}}
    \hfill
    \subfloat[]{\includegraphics[width=.445\linewidth]{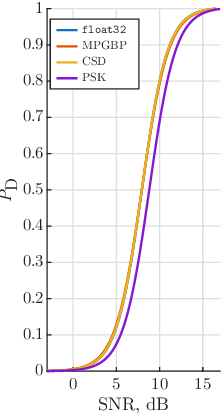}\label{fig:pd-snr}}
    \caption{Probability of detection, $P_{\textrm{D}}$, in terms of (a) false alarm rate, $P_{\textrm{FA}}$, at $5~\textrm{dB}$ SNR, and (b) SNR for $P_{\textrm{FA}}=10^{-3}$. Considering approximations with comparable complexity, MPGBP using $L_{\textrm{max}} = 420$, 3-bit CSD, and 16-PSK with CORDIC using $\theta=1$, and the non approximated sequences \texttt{float32}.}
    \label{fig:detect-prob}
    \vspace{-2.5ex}
\end{figure}

Lastly, we assess the correlation detection performance in Fig.~\ref{fig:detect-prob}; (a) presents the receiver operating characteristic (ROC) curve at a $5~\textrm{dB}$ signal-to-noise ratio (SNR), and (b) the probability of detection, $P_{\textrm{D}}$, at a constant false alarm rate, $P_{\textrm{FA}}=10^{-3}$. The simulation uses $384143$ samples for the Monte Carlo process. We consider approximations with comparable complexity, $A=839$, $1116$, and $1677$ for MPGBP ($L_{\textrm{max}}=420$), $3$-bit CSD, and 16-PSK ($\theta=1$), respectively. Although CSD and PSK+CORDIC use the minimum feasible complexity, MPGBP still uses fewer additions and could be further reduced. However, this approximation is interesting because it reduces PAPR. Fig.~\ref{fig:roc} shows that the MPGBP and CSD have an ROC similar to \texttt{float32}; however, PSK with one CORDIC iteration cannot attain the same performance level. Fig.~\ref{fig:pd-snr} shows the same behavior across different SNR values at $P_{\textrm{FA}} = 10^{-3}$.
When we compare the modified and the original MPGBP, we observe that our proposition is more robust to the parameter $L$. Considering $L=28$ and $k_{\textrm{max}}=6$, the original MPGBP results in budget errors in the order of $95\%$ of $L$ for $L_{\textrm{max}}$ varying from $84$ to $1846$ with steps of $84$, while the proposed algorithm achieves the best feasible solution given $L_{\textrm{max}}$ and $k_{\textrm{max}}$, i.e., no budget errors. To achieve similar results using both algorithms, the original algorithm requires fine-tuning.

\section{conclusion}\label{sec:conclusion}

This work shows that MP-based SOPOT approximations of ZC sequences enable a more cost-effective approach to low-complexity computation than the PSK+CORDIC and CSD approaches. Results show that, in terms of the number of bits required to store the sequences, the MPGBP approximation falls between them, with PSK being less demanding. However, the MPGBP outperforms both methods in the number of additions required to perform the inner product. The MPGBP approximation achieves similar matched-filter detection performance as \texttt{float32}. Compared to the original algorithm our modified MPGBP improves budget management of the number of SPTs and provides robustness to the algorithm's approximation parameters.

This work considers encoding the sequences using a fixed number of bits per element. Although this approach suits low-complexity decoders, MPGBP is known for achieving sparse vector approximations~\cite{da2014fir,da2021adaptive,coelho2025multiplierless}. Future work would benefit from exploring such sparsity in compression.

\appendices
\section{Quantization Function} \label{app:qtz-fun}

Finding $k_m$ that best approximates $2^{-k_m} = Q\left( \frac{{\bf r}_m^T\boldsymbol{\phi}_{p_m}}{L_{m}} \right) \approx \frac{{\bf r}_m^T\boldsymbol{\phi}_{p_m}}{L_{m}}$ becomes a quantization problem, and can be solved as
\begin{align}
    2^{-k_m} \frac{(1+2^{-1})}{2} &\leq \frac{{\bf r}_m^T\boldsymbol{\phi}_{p_m}}{L_{m}} < 2^{-k_m+1}\frac{(1 + 2^{-1})}{2} \nonumber \\
    k_m &\geq -\log_2 \left( \frac{4{\bf r}_m^T\boldsymbol{\phi}_{p_m}}{3L_{m}}\right) > k_m - 1,
\end{align}
yielding, \vspace{-3ex}
\begin{align}
    k_m = \left\lceil -\log_2 \left( \frac{4{\bf r}_m^T\boldsymbol{\phi}_{p_m}}{3L_{m}} \right) \right\rceil.
\end{align}

\section{Vector Selection} \label{app:vect-select}

At the $m$-th iteration, we select, from the dictionary, a vector $\boldsymbol{\phi}_p$ that maximizes the inner product. This is expressed as,
\begin{align}
    p_m = \arg \max_{p \in \Gamma} ~ \left\langle {\bf r}_m, \frac{\boldsymbol{\phi}_p}{\sqrt{L_{m}}} \right\rangle. \label{eq:seek-element}
\end{align}
Since $\boldsymbol{\phi}_p \in \mathcal{M}^N$, with $\mathcal{M} \in \{-1, 0, 1\}$, and $L_{m} = ||\boldsymbol{\phi}_p||_0$, the maximization is as follows: let $\mathcal{N}=\{0, 1, \dots, N-1\}$ be the set of indexes of ${\bf r}_m$, we define a set $\mathcal{L} \subset \mathcal{N}$ that selects the $L$ entries of ${\bf r}_m$ with greatest absolute values; then, the dictionary vector is build as
\begin{align}
    \phi_{p_m}(n) = \begin{cases}
        \textrm{sign}(r_m(n)), &\quad n\in\mathcal{L} \\
        0, &\quad \textrm{otherwise}
    \end{cases}.
\end{align}

\bibliographystyle{IEEEtran}
\bibliography{ref}

\end{document}